\documentclass[12pt]{article}

\usepackage{authblk}       % for affiliations
\usepackage{algorithm}
\usepackage{algorithmic}

\usepackage{float}
\usepackage{graphicx}

\usepackage{amsmath}
\usepackage{url}
\title{Versatile 3D reconstruction framework for hard X-ray grazing incidence imaging of nanostructures}

\author[1]{Luke Besley}
\author[1]{P. S. J{\o}rgensen}
\author[2]{A. Diaz}
\author[3]{C. Detlefs}
\author[1]{S. De Angelis}
\author[2]{M. Carlsen}
\author[4]{B. Chang}
\author[4]{C. Silvestre}
\author[1,*]{J. W. Andreasen}

\affil[1]{Technical University of Denmark, DTU Energy, 310, Fysikvej, DK-2800 Kgs. Lyngby, Denmark}
\affil[2]{PSI Center for Photon Science, Paul Scherrer Institut, 111, Forschungsstrasse, 5232 Villigen PSI, Switzerland}
\affil[3]{European Synchrotron Radiation Facility, 71, avenue des Martyrs, CS 40220, 38043 Grenoble Cedex 9, France}
\affil[4]{Technical University of Denmark, DTU Nanolab, 347, Oersteds Plads, DK-2800 Kgs. Lyngby, Denmark}
\affil[*]{Corresponding author: jewa@dtu.dk}

\date{} % empty = no date printed

\begin{document}
\maketitle

\begin{abstract}

Coherent imaging techniques such as ptychography offer powerful capabilities for 3D resolution of nanoscale structures. By application in grazing incidence, such techniques may achieve exceptional surface sensitivity as demonstrated by grazing incidence small angle scattering. This requires however an extension of the conventional analysis based on the Distorted Wave Born Approximation which is typically limited to stratified-layer models and statistical descriptions of in-plane structures. The prevailing implementations of reconstruction algorithms for ptychography based on the projection approximation fails to capture the significant multiple scattering that occurs in grazing incidence. We present a ptychographic reconstruction framework that replaces the single-scattering model with a multislice wave-propagation formalism tailored to grazing incidence. The framework supports simultaneous phase retrieval and reconstruction, and can incorporate multiple incidence angles, multiple rotation angles, and flexible experimental geometries into a single inversion. Reconstructions can be initialized from a random guess without strong structural priors, enabling the recovery of complex surface and near-surface nanostructures. This reconstruction framework is applied to both experimental and simulated datasets, demonstrating its versatility.

\end{abstract}

%%%%%%%%%%%%%%%%%%%%%%%%%%  body  %%%%%%%%%%%%%%%%%%%%%%%%%%
\section{Introduction}

Grazing incidence small-angle X-ray and neutron scattering techniques (GISAXS/GISANS) are used for their excellent surface sensitivity\cite{Nielsen2011}. With a low penetration depth into the substrate, the techniques are well-suited for studying the surface and near-surface regions of materials.

GISAXS and GISANS are most commonly interpreted using the Distorted Wave Born Approximation (DWBA)\cite{Sinha1988,Parratt1954,Holy1994}. Several well-established software packages such as BornAgain \cite{Pospelov2020} and FitGISAXS \cite{Babonneau2010} which utilize the DWBA are available. These methods treat the beam as incoherent and the sample as a set of stratified layers parallel to the surface, which is often referred to as stacking of multiple layers akin to the Parratt formalism \cite{Parratt1954} for modeling the reflectivity curve of multilayer structures.

DWBA-based approaches usually account for in-plane structures in a statistical sense, such as through surface roughness or lateral correlation lengths. However, recent developments have extended such approaches to resolve explicit in-plane structures. These approaches are based on formulations assuming a multilayer stack along $z$ and require known form factors\cite{Yang23,Chu2023,Jiang2011,Sun2012}. 

In this work we present an approach to analyzing coherent GISAXS data based on multislicing\cite{Cowley1957,Ishizuka1977,Hare1994,Thibault2006} and ptychography\cite{Hoppe1969,Hoppe1969B,Hoppe1969C,Rodenburg2004,Pfeiffer2018}. The approach differs fundamentally from the commonly used interpretation based on the DWBA by considering slices perpendicular to the surface, rather than layers parallel to the surface. Below, we introduce the concept of multislicing and how it is applied to the grazing incidence geometry, present a first experimental implementation, demonstrate further capabilities in simulated experiments, and finally discuss the advantages and limitations of the approach. 

Coherent diffraction imaging (CDI) techniques such as ptychography originated in the context of imaging thin objects in transmission\cite{Miao2012}. These methods rely on wave optics and are described using the framework of Fourier optics, which is based on the paraxial approximation, i.e.~that all scattering angles with respect to the optical axis are assumed to be small\cite{goodman2005introduction}. This is generally the case for X-rays, with the notable exception of Bragg diffraction.

 Ptychography is a technique to obtain the complex object function of the sample from intensity measurements. This requires an iterative reconstruction of the phase. To obtain the necessary oversampling, the positioning of the sample is varied in space with respect to the illuminating wave function (probe) in a series of overlapping positions. In most cases, the projection approximation is used, i.e.~the sample's index of refraction is projected into a single plane\cite{Nugent2010,Suzuki2014}. This does not allow for multiple interactions of the wave field with the sample; furthermore the approximation starts to break down when significant propagation happens within the sample (this is known as the depth of field limitation).

\begin{figure}[H]
    \centering
    \includegraphics[width=\linewidth]{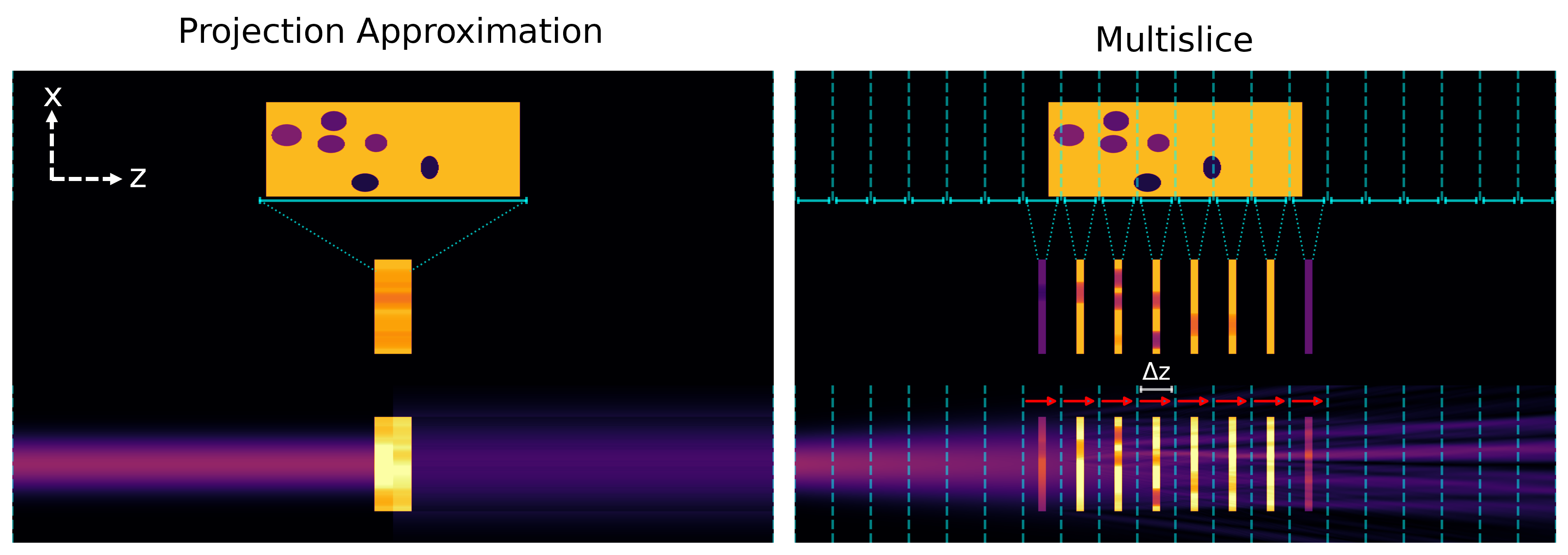}
    \caption{ In the projection approximation (left), the sample’s transmission function is projected onto a single plane. In multislicing (right), the sample is divided into several slices perpendicular to the optical axis ($z$), and each of these slices is projected onto a separate plane. The wave field is then propagated from plane to plane, allowing for propagation effects to arise within the sample volume and for multiple interactions of the wave field with the sample.}
    \label{fig:ProjectionApproximation}
\end{figure}

To overcome these limitations, multislicing was introduced \cite{Cowley1957,Ishizuka1977,Hare1994,Godden16,Tsai:16,li2018}.
Here, the sample is divided into multiple slices perpendicular to the optical axis, and each slice is projected onto a plane. 
The wave field is then propagated from plane to plane, allowing for propagation effects within the sample and for multiple interactions of the wave field with the sample, see Fig.~\ref{fig:ProjectionApproximation}.

\begin{figure}[H]
    \centering
    \includegraphics[width=\linewidth]{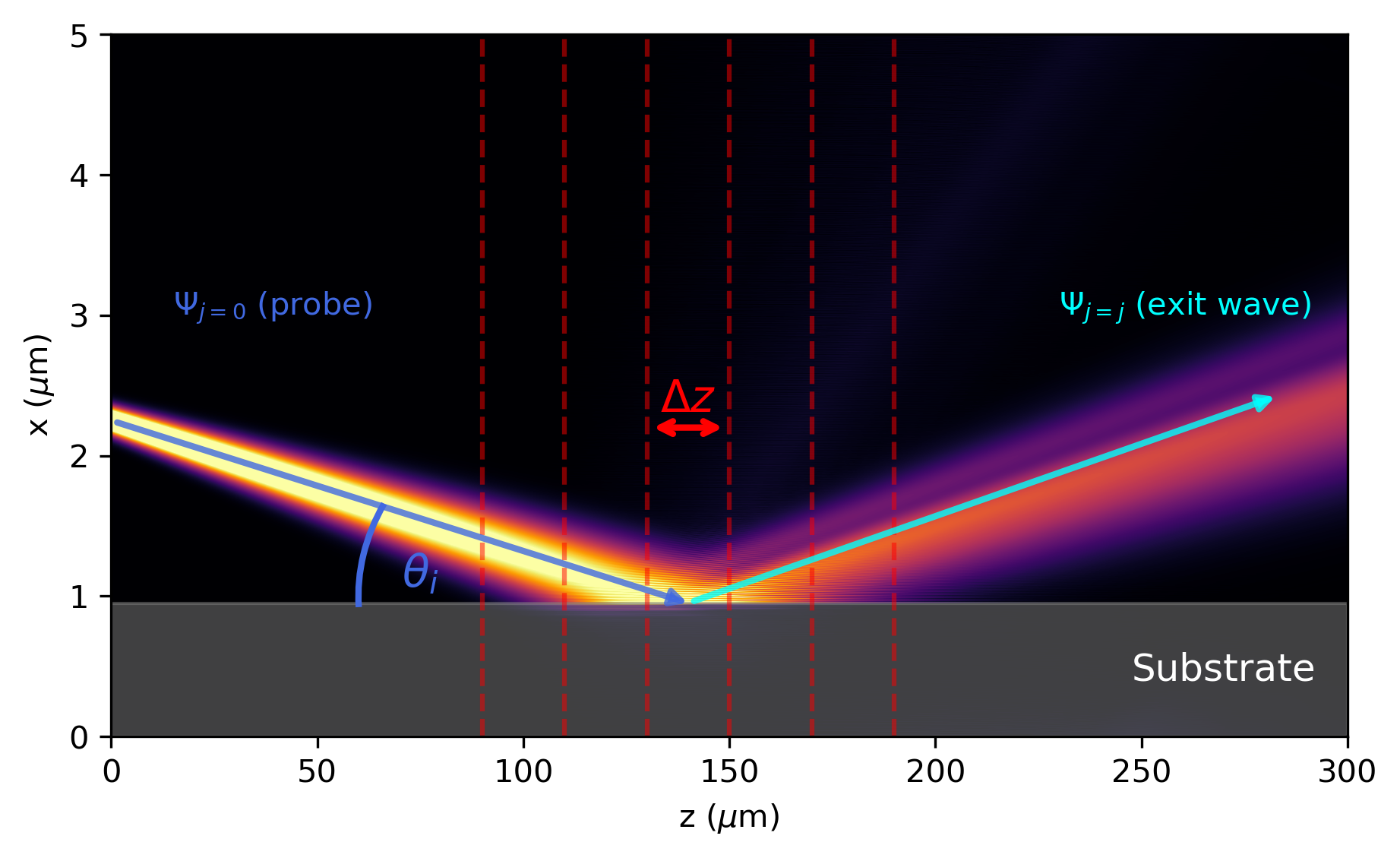}
    \caption{The grazing incidence geometry is simulated by aligning the optical axis parallel to the substrate surface, such that the slices are perpendicular to the substrate. The angle of incidence is then applied to the incident wave field in the form of a constant phase gradient. Reflection and multiple scattering arise naturally. Here $\theta_i$ is the angle of incidence, while $\psi_j$ refers to the $j_{th}$ wave in the simulation. $\Delta z$ is the slice thickness.}
    \label{fig:grazinggeometry}
\end{figure}

In our grazing incidence configuration, the angle of incidence is modeled as a constant phase gradient in the probe function, see Fig.~\ref{fig:grazinggeometry}. The coordinate system is such that $x$ is normal to the substrate, the wavefront (and slicing direction) propagates along $z$, with $y$ being in the plane of the substrate and transverse to the beam propagation direction. %Reflection of a substrate and multiple scattering within a multilayer stack are naturally taken into account by this model \cite{multislice}, also see supplementary information.

% (some text on the development and recent use of multislicing)
In transmission geometry, multislice ptychography has seen an uptake in recent years\cite{kahnt2021,li2018,Tsai:16,Du2021,Shimomura:18}. Typically, the number of slices used is small, less than 10, and is used to overcome depth of field resolution limits from standard ptychography. Multislice has been successfully shown in Li. et al \cite{Li2017} to reproduce grazing incidence reflectivity of X-rays interacting with a 3D distribution of refractive indices, but typically requires a much larger ($10^2$ to $10^3$) number of slices to accurately capture a fully reflected wavefront.

Recent works have seen an uptake in the use of automatic differentiation for inverse problems such as image reconstruction\cite{Du2021} and phase retrieval\cite{Jurling2014} to handle gradient computation of the loss function for complex forward models. With the advantage of GPU acceleration through automatic differentiation packages such as Pytorch, it has become feasible to incorporate multislice as a differentiable forward model to solve real-space structures from coherent diffraction data for reflection geometries\cite{Myint2023,Jorgensen:24,Sung2025}.

\begin{figure}[H]
  \centering \includegraphics[width=0.75\columnwidth]{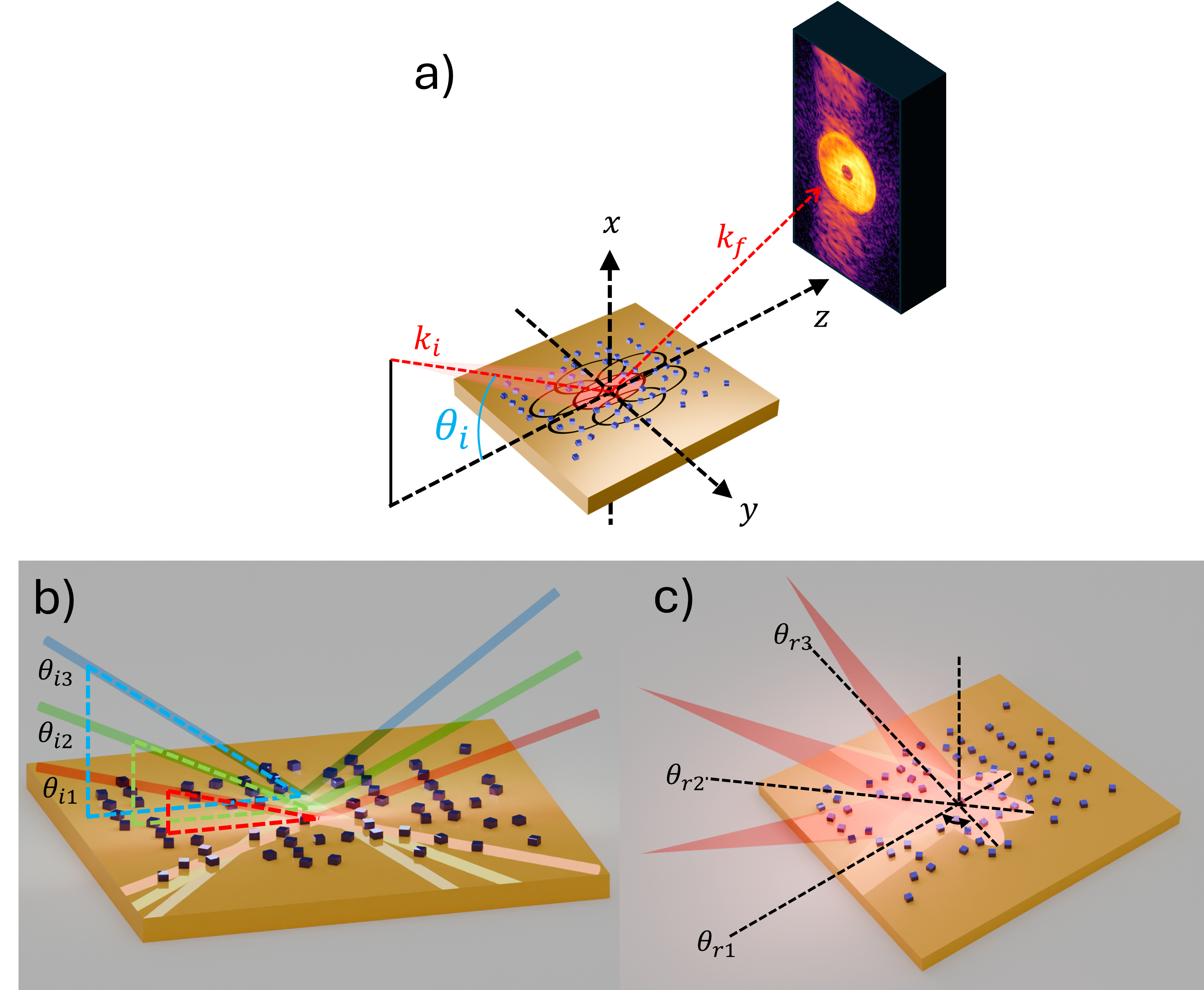}
    \caption{\label{fig:allgeometries} a) A typical experimental setup for grazing incidence X-ray ptychography. Black circles indicate overlapping scan positions which are acquired as the sample is translated along the surface. b) a schematic illustrating the acquisition of multiple incidence angles used in a single reconstruction. c) Schematic illustrating another imaging mode akin to laminography, where different scans are acquired as the sample is rotated about its surface normal. Note that in b) and c) only a single beam footprint is shown for clarity.}  
  \end{figure}

In this work, we apply the multislice formalism to the analysis of coherent GISAXS data, specifically to ptychographic reconstruction.
While simulations of GISAXS using multislice techniques have been shown \cite{Myint2023,Venkatakrishnan2020,Sung2025}, the ability to determine the real-space structure of the scattering sample from a random initial guess represents as significant step forward. We incorporate this formalism into a new framework suited for grazing incidence ptychographic reconstruction with a variety of flexible scanning geometries as shown in Fig.~\ref{fig:allgeometries}. We demonstrate a faithful 3D reconstruction of nanostructures on a surface from experimental data imaged in grazing incidence, with very high resolution along $x$ (on the length scale of $10^0-10^1$ nm), intermediate resolution along $y$ ( $10^1-10^2$ nm) and lower resolution along $z$ ($10^2-10^3$ nm), starting from a random initial guess, using multislice with automatic differentiation, performing phase retrieval and 3D reconstruction in a single algorithm. By using multislice as a forward model, a ptychographic data set can be directly reconstructed into 3D, where the only required priors are experimental conditions required for conventional ptychography, i.e. wavelength, an accurate description of the wave front, and scan positions.

The framework is provided as an open source Python package, PyGRAPES (\textbf{Py}torch \textbf{GRA}zing incidence \textbf{P}tychographic \textbf{E}ngine for \textbf{S}urfaces and nanostructures), with working example notebooks, available on github\cite{PyGRAPES_2025}.

\section{Method}
\subsection{Projection approximation}

In the projection approximation, the sample is represented by a single \emph{object} function $O(x, y)$ where the sample's refractive index is integrated along the optical axis,
\begin{equation}
    O(x, y)	=\exp\left[i k \int n(x, y, z) \,\mathrm{d}z \right].
\end{equation}
	
Here $k = 2\pi/\lambda$, $\lambda$ is the X-ray wavelength, and $n(x, y, z) = \delta(x,y,z) + i \beta(x,y,z)$ is the spatially varying complex refractive index of the sample. The optical axis is along $z$. The exit wave function just downstream of the sample is then given by
\begin{equation}
\psi(x,y) =	O(x, y) P(x, y)
\end{equation}
where $P(x, y)$ is the illuminating wave function just upstream of the sample, often referred to as the \emph{probe}, see Fig.~\ref{fig:grazinggeometry}. 
Finally, the exit wave function is propagated to the detection plane, where the detected intensity is given by the squared magnitude of the propagated exit wave. In the far field approximation, the propagation reduces to a Fourier transform.

\begin{align}
	I_m(x, y)
	&=
	\left| {\cal P}(z_d) \psi(x,y) \right|^2
	\\
	&=
	\left| {\cal P}(z_d) O(x, y) P(x-x_m, y-y_m) \right|^2
	\\
	&=
	\left| {\cal F}\left[O(x, y) P(x-x_m, y-y_m)\right] \right|^2,
\end{align}
where ${\cal P}(z_d)$ is the Fourier optics propagator through distance $z_d$, and $z_d$ is the distance from the sample to the detector.
This intensity is recorded for a series of probe beam positions $(x_m, y_m)$.

\subsection{Multislicing}

In multislicing, the sample is divided into several slices perpendicular to the optical axis, and the beam is propagated from slice to slice, see Fig.~\ref{fig:grazinggeometry}.
Let the slices be spaced by $\Delta z$. The sample is then represented by a stack of object functions $O_j(x,y)$ with

\begin{align}
	O_j(x, y)
	&=
	\exp\left[ i k \int_{(j-1)\Delta z}^{j \Delta_z} n(x, y, z) \,\mathrm{d} z \right].
    \label{equation:wave_interact}
\end{align}

The wave function is iteratively propagated from slice to slice,
\begin{align}
	\psi_j(x, y)
	&=
	{\cal P}(\Delta z) O_j(x, y) \psi_{j-1}(x, y),
    \label{eqn:wave_propagate}
\end{align}
with $\psi_0(x, y) = P(x - x_m, y-y_m)$ being the probe function.
The exit wave function is then given by $\psi_J(x,y)$, where $J=Z/\Delta z$ is the number of slices for a sample with thickness $Z$. This is again propagated to the far field detector.

\subsection{Reconstruction Methods}
The reconstruction volume is discretized into a $N_x \times N_y \times N_z $ grid, with separate voxel resolutions of $\Delta x, \Delta y, \Delta z$ for each dimension. In the case of grazing incidence X-ray ptychography, the resolution and field of view is highly anisotropic and varies greatly between axes. Therefore, an appropriate reconstruction volume size and voxel resolution must be chosen for each axis. A typical volume size is on the order of $N_x$ = 2000, $N_y$ = 500 and $N_z$ = 1000 voxels, with voxel sizes of $\Delta x =$ 5~nm, $\Delta y$= 40~nm, $\Delta z$= 500~nm. 

The input probe is a model Fresnel zone plate (FZP) based on the optics used for experimental data as described in \ref{section:experimental_methods}. The model probe is generated using a modified version of the ptychoshelves code for generating a model FZP\cite{Wakonig2020}. As in conventional ptychography, a previously optimized probe can be used as the starting guess. One probe is used per scan, comprised of a Fermat spiral of approximately 300-500 points per scan. 

The probe interacts with each slice of the volume according to eqn. \ref{equation:wave_interact}, and is then propagated according to eqn. \ref{eqn:wave_propagate}. The propagator is based on the Fresnel approximation and adapted from the implementation used in the Adorym package\cite{Du2021}. This process is repeated for all slices of the volume, and the final exit wave after $N_z$ slices ($\psi_{j=N_z}$), is Fourier transformed and the square amplitude is taken. This output, $G_{guess}$, is compared to the experimental data $G_{data}$ according to conventional MSE loss:
\begin{equation}
    \mathrm{Loss}\ = \frac{1}{N_{p}}\sum_{i=1} ^{N_{p}}(\mathrm{G_{guess}}-\mathrm{G_{data}})^2+\tau+\gamma
\end{equation}
Where $N_{p}$ is the number of pixels in $G$, $\gamma$ is an estimate of noise, and $\tau$ denotes additional regularization terms incorporated into the loss function, such as an anisotropic total variation regularization (see supplementary information for more details). Following this, backward propagation of the gradients of the loss function are calculated for a single forward simulation, and stored into a buffer tensor for the local area of the volume which was illuminated for a given scan point (in this context, gradient backpropagation refers to the reverse-mode calculation of the loss-function derivatives via automatic differentiation). This is an important requirement to avoid excessively high memory requirements, as only one volume used in the forward simulation, which typically requires 10 to 20 GB of VRAM, can be stored at one time on a conventional, commercially available GPU. This can of course be parallelized if more memory storage is available. When gradients for all scan positions are calculated, the gradients stored in the buffer tensor are accessed. From here, the volume is finally updated and a single iteration is complete.
A brief outline of the algorithm is shown in algorithm \ref{algorithm:multislice}. The addition of more incidence angles or rotation angles simply adds to the number of forward simulations.

\begin{algorithm}
\caption{Multislice Optimization Procedure}
\begin{algorithmic}[1]
\REQUIRE Number of iterations $N$, diffraction patterns $M$, scan positions $p$, volume $\mathbf{v}$,
\FOR{$i = 1$ to $N$}
    \FOR{$ii = 1$ to $M$}
        \STATE Zero gradients
        \STATE Select sub-volume $\mathbf{v}_{ii}$ from sample with scan position $p_{ii}$
        \STATE Compute diffraction pattern output from multislice forward simulation of $\mathbf{v}_{ii}$
        \STATE Compute MSE loss for $\mathbf{v}_{ii}$ and diffraction pattern $M_{ii}$
        \STATE Compute gradients via backward propagation of loss function
        \STATE Store and accumulate gradients in buffer for $\mathbf{v}_{ii}$ at position $p_{ii}$
    \ENDFOR
    \STATE Transfer all gradients to $\mathbf{v}$
    \STATE Perform optimization step
\ENDFOR
\end{algorithmic}
\label{algorithm:multislice}
\end{algorithm}

\subsubsection{Constraints}
It is important to discuss how the full 3-dimensional structure is formed. The volume is constructed in such a way that the structure must be a solid, continuous structure built upward from the substrate. The sample is initially encoded as a 2D matrix (in Pytorch referred to as a tensor), where each value corresponds to the height of the 3rd dimension along $x$ at position $(y,z)$. During the forward pass, the sample is converted to 3D by repeating the structure a given number of times along its 3rd dimension (normal to the substrate) for each 2D voxel. The 3D structure is passed into the forward model, and for gradient calculation, all refractive index changes are therefore converted back to height. The volume can also be optimized purely as a 3D volume, however a conformance to a height map representation achieves better convergence.  %This operation must be differentiable, and so is implemented using native pytorch repeat and expand functions. A material constraint is also implemented. 
In the samples imaged experimentally, the structure is known to be entirely Au, and we can reduce the reconstruction to a 2-phase system, i.e. the reconstructed volume must consist of either Au (the substrate and structure material in this case), or air, exclusively. This allows us to constrain our possible solution space, and to implement a formula for partial voxel mixing, which allows us to surpass the vertical resolution determined by the voxel height. By implementing this 2-phase constraint, we can reformulate equation \ref{equation:wave_interact} to be as follows:

\begin{equation}
\label{equation:wave_interact_partial_voxel}
    \psi_j* = \psi_j \times [c \cdot \exp[ik(\delta_{\text{Au}}+i\beta_{\text{Au}})\Delta z] + (1-c)\cdot\exp[ik(\delta_{\text{Air}}+i\beta_{\text{Air}})\Delta z]]
\end{equation}

where $c$ is a value between 0 and 1 corresponding to the volume fraction of Au (or any substrate/structure material) within a given voxel, as described and verified by Li et. al\cite{Li2017} in their work using a multislice algorithm. The framework therefore reconstructs a 3D  volume where each voxel has a complex-valued $\delta +i\beta$, rather than a phase and amplitude object, and equation \ref{equation:wave_interact_partial_voxel} allows the object function to remain continuous within a single voxel, which is required for automatic differentiation. 

By implementing equation \ref{equation:wave_interact_partial_voxel} into our forward model as the wave interaction step, we can restrict the solution space of each voxel to be between 0 and 1, and more effectively implement this constraint that ensures solid physical continuity from the substrate. These constraints additionally have the effect that the sample must be composed of a mixture of known refractive indices.% More complex systems can be implemented by further increasing the complexity of the forward model. 

Position refinement can be carried out by applying the 2D Fourier shift theorem \cite{Lim1990} of the incident probe within the simulation volume prior to performing multislice and refining the shift as a model parameter. This method has been successfully implemented in the Adorym package for position refinement\cite{Du2021}.

Various optimizer packages exist within PyTorch. The Adaptive Moment estimation (ADAM)  \cite{kingma2017adam} is particularly popular in use. For the optimization of these reconstructed volumes, we have found that PyTorch's standard stochastic gradient descent (SGD) optimizer performs better in fewer iterations. Thus, SGD  was used for optimization of the volume, while other optimizable parameters, such as the probe, and position refinement, are done with ADAM. The final reconstruction uses a pre-optimized probe as its initial guess. In the case of a reconstruction with multiple incidence angles,  offsets for each incidence angle are manually identified and added to the initial guess of motor positions.

In the case of a reconstruction for scans that are are acquired via rotation around the surface normal, a transformation of the volume occurs before beginning the multislice, however Algorithm \ref{algorithm:multislice} still remains the same. Firstly, the transverse and longditudinal resolution are both set to be the transverse resolution (as typically $\Delta z >> \Delta y$), i.e. $\Delta z_{final} = \Delta y $. For each rotational angle $\theta_r$, The 2D representation of the volume is first rotated to $\theta_r$ using PyTorch's in-built grid sample and affine transformations \cite{pytorch_affine_grid}, and then using a 2D convolution operator, downsampled along the z-dimension to the initial $\Delta z_{init}$ that is used in the forward model as previously described. During the backward calculation, the volume is de-convoluted so that $\Delta z = \Delta y$ again. This is done to allow the multislice volume to fit into memory. The final reconstructed volume has high resolution $\Delta y = \Delta z$. $\Delta x$ remains unchanged.

\subsection{Experimental Methods}
\label{section:experimental_methods}
Experimental data were acquired at the cSAXS beamline of the Swiss Light Source (SLS) at the Paul Scherrer Institute in Villigen, Switzerland, using a photon energy of 6.2~keV. Coherent illumination on the sample was defined by a FZP made of Au, fabricated by the X-ray nano-optics group at the Paul Scherrer Institute \cite{Gorelick2011}. The FZP had a diameter of 220$~\mu$m and 90~nm outer-most zone width, resulting in a focal length of 99~mm and a focal depth of $\pm80~\mu$m. A Pilatus 2M detector with a pixel size of 172$~\mu$m was used for collection of diffraction data at a sample to detector distance of 7.36~m. At the X-ray photon energy of 6.2 keV, the critical angle $(\theta_{c})$ for Au is $\theta_c \approx 0.72^\circ$. The reconstructions in this work were taken at incidence angles $(\theta_{i})$ of 0.5, 0.6, and 0.8$^\circ$. For further details about experimental methods we refer to previous work\cite{Jorgensen:24}.

\section{Results}

\subsection{Reflectivity from a multilayer stack}

\begin{figure}[H]
  \centering \includegraphics[width=\columnwidth]{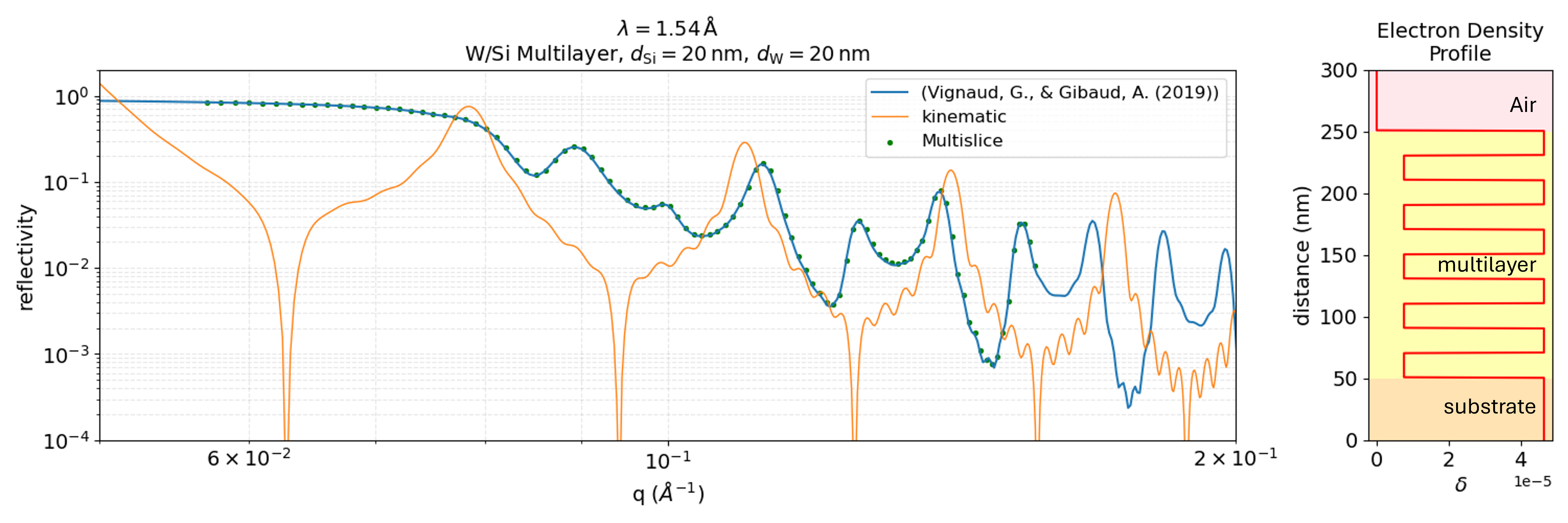}
    \caption{\label{fig:Reflectivitycurve} Reflectivity curve of W/Si multilayer at $\lambda = 1.54Å$. Kinematic curve calculated as in Nielsen \cite{Nielsen2011}. Dynamical curve calculated using slab-model approach using the Abeles matrix method, as per Vignaud and Gibaud (2019)\cite{Vignaud2019REFLEX}}  
  \end{figure}

Multiple scattering (dynamical theory) is known to strongly affect the reflectivity of multilayers. We have therefore chosen a multilayer stack as a test case to compare our multislicing simulations to well-known reference approaches. Fig.~\ref{fig:Reflectivitycurve} shows the simulated reflectivity curves generated for a 5-layer stack of alternating W/Si multilayers, with $d_{Si}=d_{W}=20$~nm, at $\lambda = 0.154$~nm. The orange curve follows the kinematic approach, as outlined in Nielsen and McMorrow \cite{Nielsen2011}, which models the reflectivity curve based on the kinematical approximation, where the incident wave field does not scatter more than once when interacting with the sample. The blue curve follows the model implemented by Vignaud \& Gibauld\cite{Vignaud2019REFLEX}, which uses the slab-model approach using the Abeles matrix method to predict the dynamical scattering effects on a reflectivity curve, which allows for a recursive model to be applied simulating the scattering for each layer upwards from the substrate.  Especially in the low-q region, kinematic approximations of reflectivity tend to fail at describing the significant multiple scattering that occurs in reflection \cite{Nielsen2011}. On the right side of fig. \ref{fig:Reflectivitycurve} the 1D electron density profile, as a function of the $\delta$ component of the complex refractive index $n$, is shown. 

The green dotted points show the reflectivity curve calculated by using our multislice approach. It can be seen that reflectivity curve predicted by multislice closely follows that of the dynamical model (blue), and therefore accurately models the complex multiple scattering phenomena that occurs in reflection. The discrepancy between the dynamical/multislice approach and the kinematic approach illustrates the need for a forward model that accurately captures multiple scattering. 

\subsection{Multiple scattering for a 2D sample}

The need to account for dynamical scattering shown in Fig.~\ref{fig:Reflectivitycurve} is well demonstrated for a 1D reflectivity stack. Of more interest in ptychography is the reconstruction of specific surface or near-surface structures. We therefore choose a test case with a specific structure to investigate the effect on a 2D sample. As shown in Jørgensen et al. \cite{Jorgensen:24}, specific structures imaged in grazing incidence with no chemical contrast can be reconstructed using the projection approximation. Structures with higher complexity, such as having both chemical and spatial variation may require a more complicated approach.

\begin{figure}[H]
  \centering \includegraphics[width=\columnwidth]{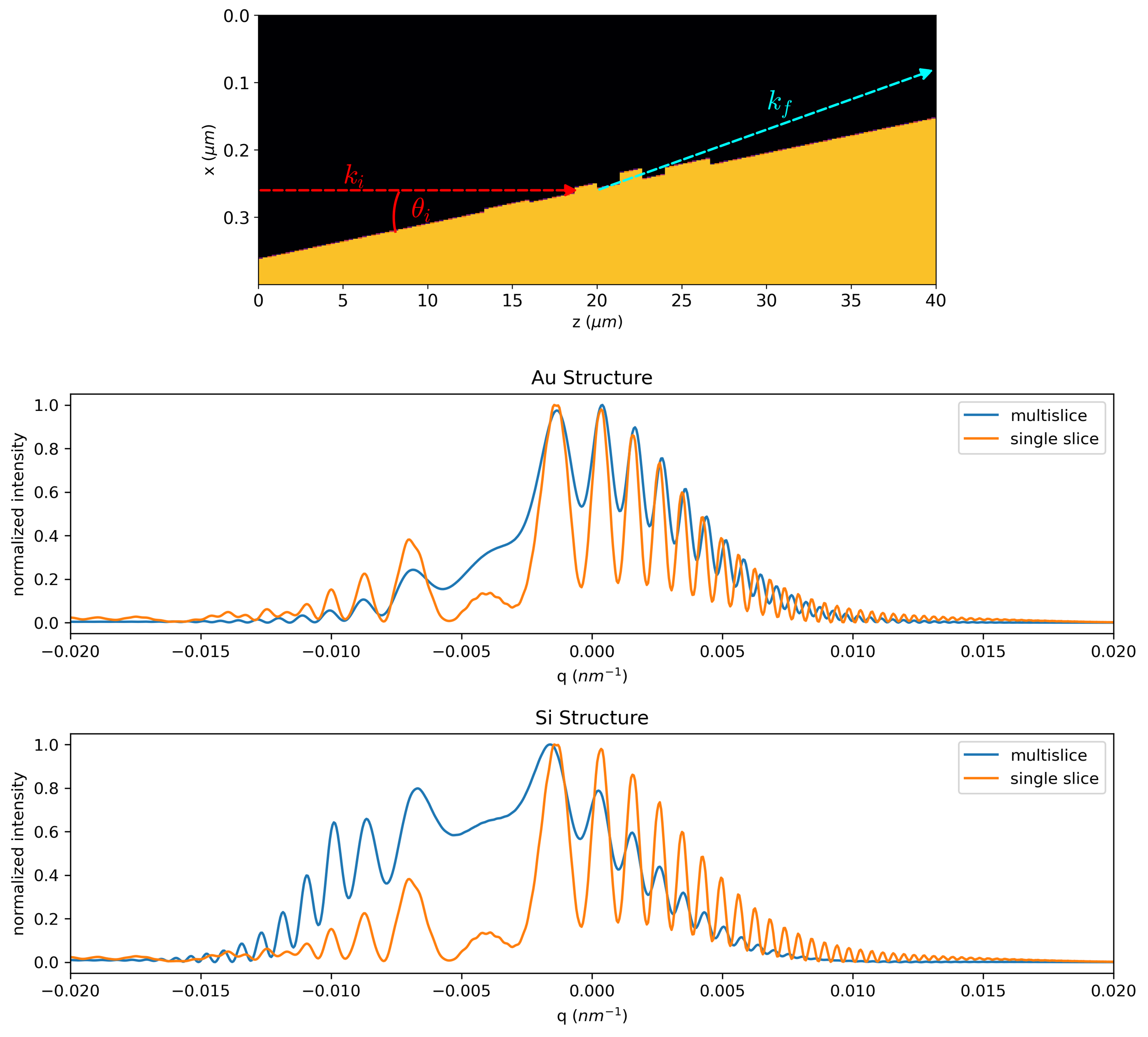}
    \caption{\label{fig:single_slice_vs_multislice} Top: a specific varying 2D structure used as the basis for both multislice and the projection approximation. The structure is entirely Au in the first case, and entirely Si in the second case, with an incidence angle of $\theta_i=0.4^\circ$ Middle: the output intensities, normalized, of the far-field exit wave, simulated via single slice and multislice with Au, showing reasonably good correlation. Bottom: The output intensities, normalized, of the far-field exit wave, simulated via single slice and multislice with Si, where multiple scattering becomes significant.}  
  \end{figure}

To verify this, we show the far-field exit wave of a simple 2D structure on a substrate, simulated as made fully by Au, and by Si. Fig.~\ref{fig:single_slice_vs_multislice} , shows a simple structure in reflection with a Gaussian beam, at $\theta_i = 0.4$. For each material, the exit wave is computed with the multislice algorithm, and then the structure is integrated into a single slice and multiplied with the probe to compute the single slice exit wave. The amplitude of the exit waves shown a reasonable correlation  between the multislice output and the single slice approximation in the case of Au, whereas the predicted amplitudes begin to differ significantly when Si is introduced, which has much weaker absorption (and hence greater transmission and multiple scattering) at this X-ray energy. This agrees with our prediction that in the case where absorption is strong, the single slice approximation may suffice, and the reconstructions will be successful, whereas when multiple scattering becomes significant, the reconstructions break down.

\subsection{Experimental Data: Verification of single incident angle reconstruction}
\begin{figure}[H]
  \centering \includegraphics[width=\columnwidth]{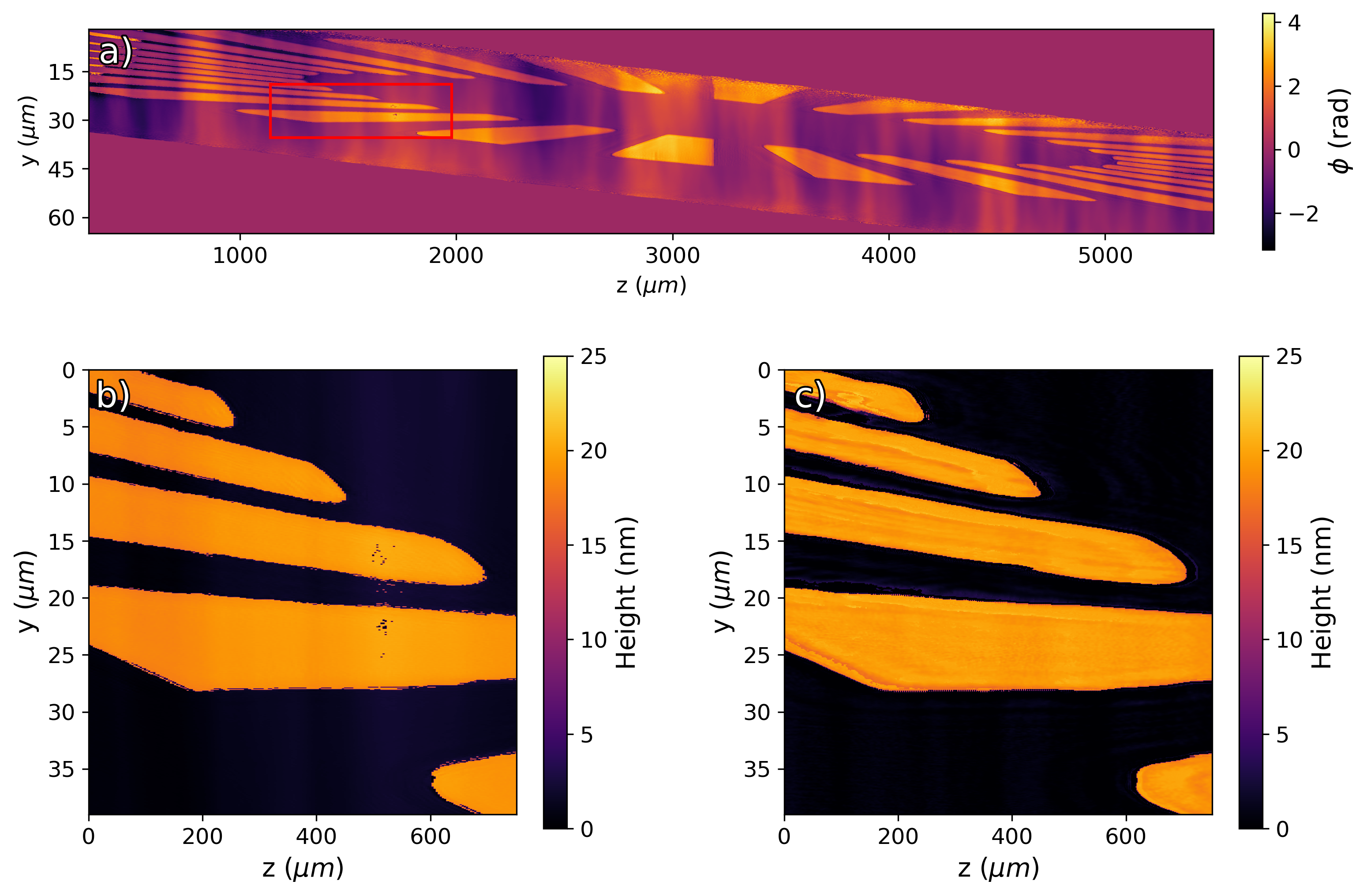}
    \caption{\label{fig:SiemensStar} Reconstructions from a single incidence angle. a) Reconstruction using the projection approximation (performed using Ptychoshelves), given in phase shifts (in radians). b) ROI of Reconstruction of a), after phase unwrapping, and converting phase shifts to height values as per the method outlined in Jørgensen et al. \cite{Jorgensen:24}. c) Reconstruction using the current framework, with the same ROI.}  
  \end{figure}

To verify the successful reconstruction of experimentally acquired data, we test the framework on a set of previously acquired data as published in Jørgensen et al. \cite{Jorgensen:24} and compare with the results for the same dataset using the reconstruction created using the ptychoshelves software. Fig.~\ref{fig:SiemensStar} a) shows the phase shift image of a large Siemens Star, taken at $\theta_i= 0.8^\circ$ with a spoke height of 20 nm, and a ROI indicated. In Fig.~\ref{fig:SiemensStar} b), the ROI of a) is shown, the phase shifts of a) have been converted into height values as per the method outlined in Jørgensen et al.\cite{Jorgensen:24}. Fig.~\ref{fig:SiemensStar} c) shows the reconstruction with the current framework. The ptychoshelves reconstruction has a longditudinal voxel size of $\Delta z = 2.04~\mu$m, whereas the reconstruction using the current framework has a voxel size of $\Delta z = 1.0~\mu$m. The shapes match well, although some slight variations in the exact shape of the spokes exist. It can be seen that the edges are slightly sharper in b), although the shape of the edges does not correlate perfectly between the two images. The mean error in height between the two images is 0.98 nm with a standard deviation of 0.7 nm. At sufficiently low spatial resolution, the projection approximation gets better. Hence, multislice and the projection approximation will give similar low-frequency information about the sample. The projection approximation breaks down at higher spatial frequencies when compared to multislice, which is consistent with what is observed in Fig.~\ref{fig:SiemensStar}.

\subsection{Experimental Data: Multiple Incidence Angles}

\begin{figure}[H]
  \centering \includegraphics[width=1\columnwidth]{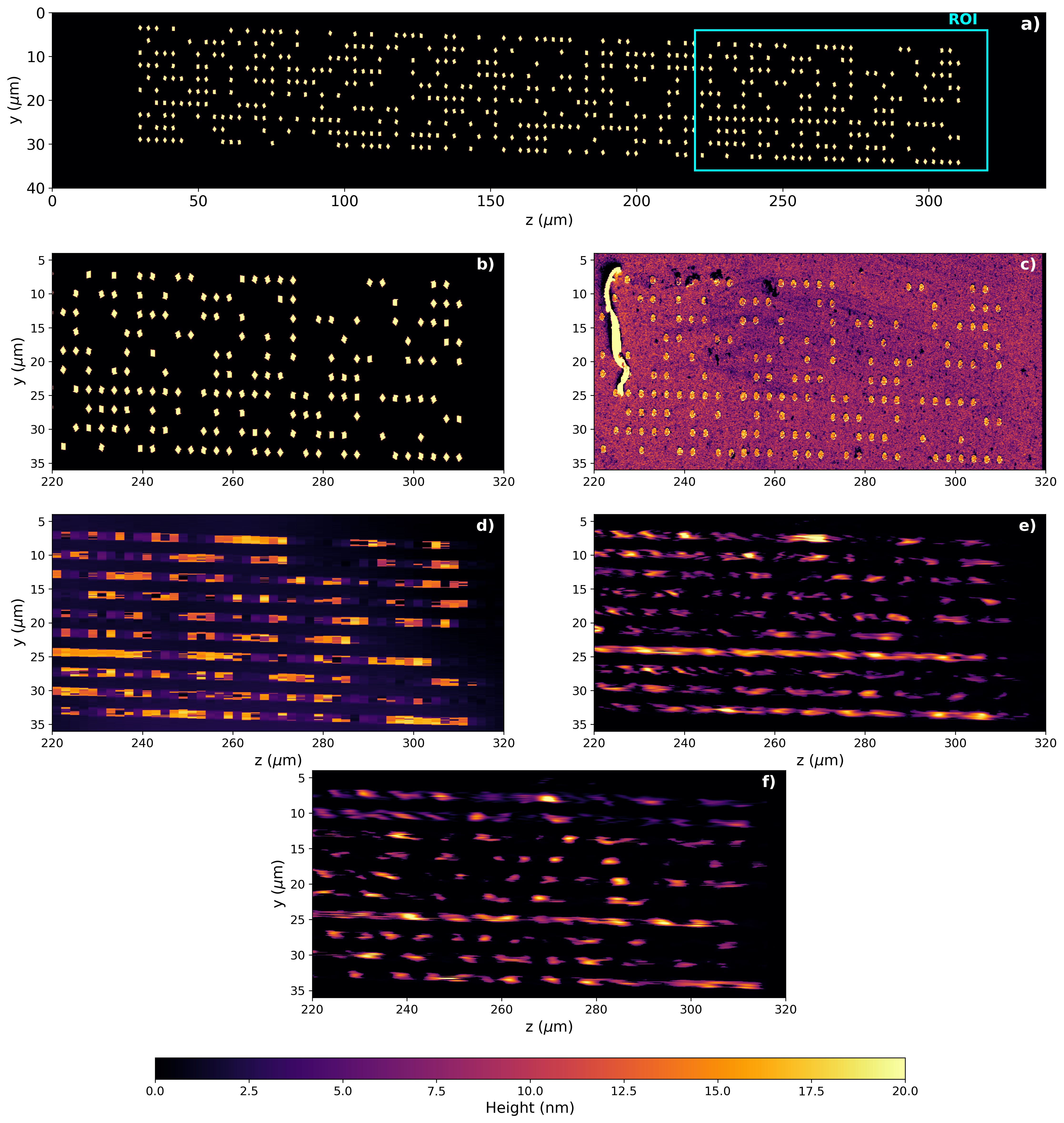}
    \caption{\label{fig:combinedscan} a) MLS Design structure, with ROI highlighted. b) ROI of the ideal design structure. c) Secondary Electron image of fabricated structure taken with scanning electron microscopy, showing roundness of squares and other defects during fabrication. d) Ptychoshelves reconstruction for $\theta_i=0.8^\circ$. e)  Height map of reconstruction with current framework for a single incidence angle ($\theta_i=0.8^\circ$). f)  Height map of reconstruction with current framework for multiple incidence angles ($\theta_i=0.6, 0.8^\circ$).}  
  \end{figure}

Fig.~\ref{fig:combinedscan} is an overview of the pattern and various reconstructions of an Au test structure on an Au substrate. Referred to as the MLS (maximum length sequence) structure, the pattern is a pseudo-random binary sequence of squares, organized into 10 rows, based on a maximum length sequence. Each square has a random rotation and design height of $20$~nm, and are spaced $2\sqrt{2}~\mu$m apart, having a length of $1\times 1~\mu$m. Squares are omitted according to the pseudo-random binary sequence. The random sequence is chosen chosen to reduce periodicity in the diffraction patterns to help aid with reconstruction. Fig.~\ref{fig:combinedscan}~a) shows the wafer design pattern of the MLS structure, with Fig.~\ref{fig:combinedscan}~b) being the ROI. Fig.~\ref{fig:combinedscan}~c) shows an SEM image acquired in secondary electron mode of the the same ROI, showing that the squares in the sequence have rounded edges (note in \ref{fig:combinedscan}~c) the color bar does not refer to height but rather detector counts). Fig.~\ref{fig:combinedscan}~d). shows the reconstruction produced from Ptychoshelves (single scattering approximation). Fig.~\ref{fig:combinedscan}~e) shows the reconstruction using the current framework, reconstructed from single incidence angle, whereas Fig.~\ref{fig:combinedscan}~f) shows the reconstruction from multiple incidence angles. When multiple incidence angles are added, the separation between squares in denser regions is better resolved as seen from Fig.~\ref{fig:combinedscan}~e) to Fig.~\ref{fig:combinedscan} f). The reconstruction does not necessarily show an overall improvement compared to the projection approximation for these samples (Fig.~\ref{fig:combinedscan}~d)), but has enhanced resolution along $\Delta z$ and shows that an improvement can be made from Fig.~\ref{fig:combinedscan}~e) by adding more incidence angles (Fig.~\ref{fig:combinedscan}~f)), thus demonstrating that the framework can successfully combine several incidence angles from experimental data. 
%This is comparable to the ptychographic reconstruction in grazing incidence where only a single incidence angle is used per reconstruction. The starting guess for all incidence angles is a substrate of Au, with a small random variation of the surface heights on the order of $\pm 0.5 $~nm as an initial guess.

It is important to consider the resolution of the reconstructed volumes. In conventional ptychography, the maximum achievable scan resolution in a given reconstruction is determined by the size of the detector window chosen, which corresponds to the highest available frequency sampled in Fourier space. In the case of multislice reconstruction, the grid size of the volume is specified in the reconstruction parameters. For the reconstructions in Fig.~\ref{fig:combinedscan}, a window size of 180 $\times$ 180 pixels is chosen. The pixel size of the ptychoshelves reconstruction with this window size is approximately 48 nm in the transverse and 3400 nm in the longditudinal direction, whereas the voxel grid of the multislice reconstruction is 40 nm in the transverse direction and 300 nm in the longditudinal direction ($\Delta z$ = 60 nm using an upsampling factor of 5 and is simulated at slices of thickness 300 nm). Although the grid size can be much smaller in multislice reconstruction along the beam direction for the same chosen window size on the detector, this does not strictly represent the true resolution of the reconstruction, which can be verified by Fourier ring correlation (FRC)\cite{Saxton1982}.
Using 1-D FRC from two reconstructions of a sample from two independent sets of data, as is described in more detail in \cite{Jorgensen:24}, for reconstruction settings of $\Delta z$ = 500 nm and $\Delta y$ = 40 nm, the calculated resolution is much poorer, 368 nm in y and 4603 nm  in z, which is approximately eight to ten times the pixel size in each direction. The two datasets collected for FRC should produce a near-identical reconstruction for each dataset, however the results from these two datasets results in appreciably different reconstructions, which at present we do not know the source of. The framework optimizes a much larger number of parameters such as position refinement, resulting in differences between two reconstructions that make perfect sub-pixel alignment much harder. Due to this difference between reconstructions, the resolution calculated by FRC is low with respect to the simulation pixel sizes $\Delta y$ and $\Delta z$. Examples of the datasets used for FRC calculation are available in supplementary information.

Resolution estimation can also be carried out by a "knife-edge" test, where an error function is fitted to a sharp edge in the image and the change between 10 and 90 percent of the intensity can be interpreted as the effective resolution \cite{Wong2019,Reimer2000,Wachulak2017}. Taking this approach from a variety of sharp edges from a variety of squares in the reconstructed MLS structure, the average resolution in $y$ was found to be 127 nm with a standard deviation of 58 nm, whereas along $z$ the average resolution is 1066 nm with a standard deviation of 231 nm, giving a resolution of approximately 2 to 4 times the simulation voxel size for both $\Delta y$ and $\Delta z$.

\subsection{Simulated data: Rotation about the surface normal}
\begin{figure}[htbp]
\centering
\includegraphics[width=\linewidth]{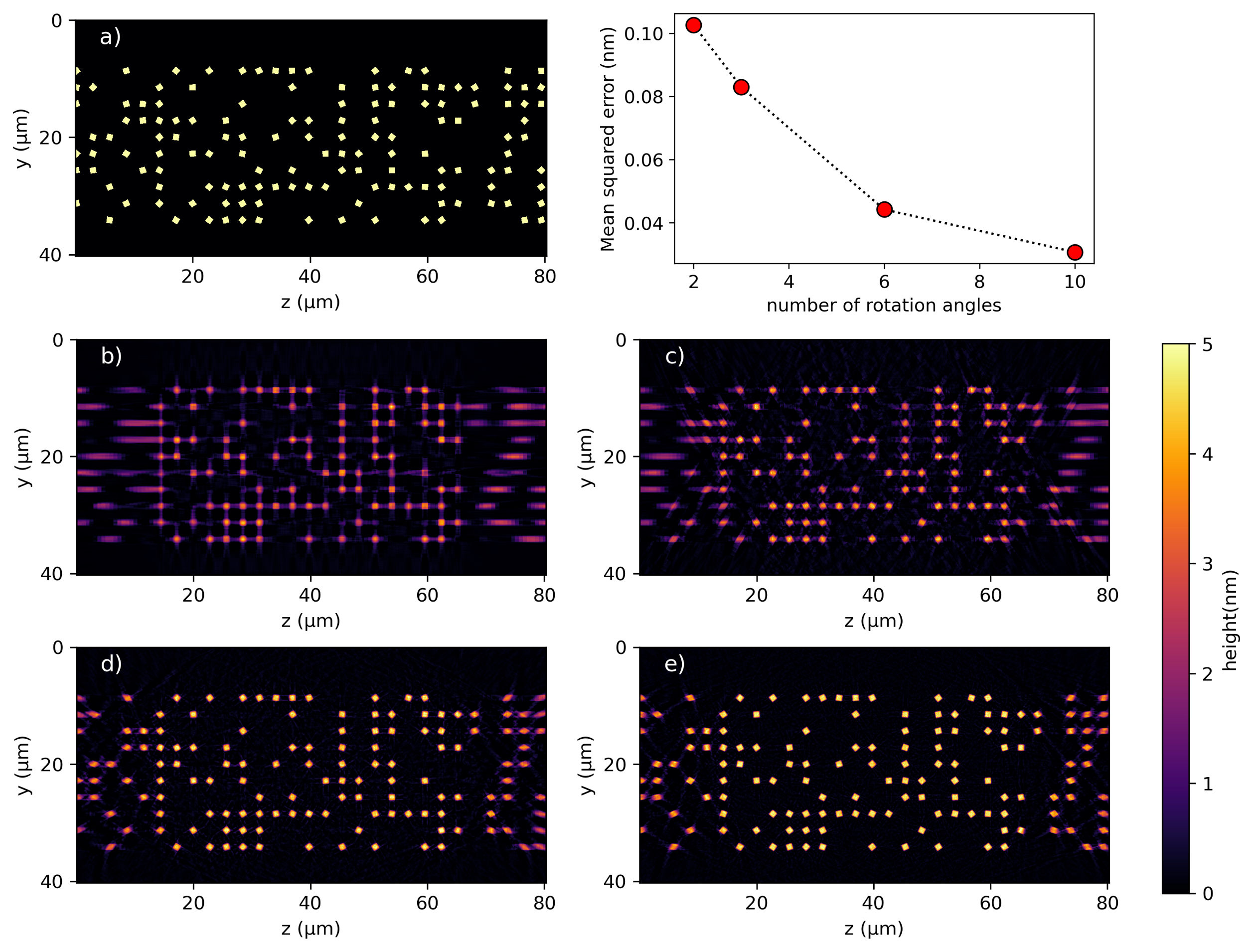}
\caption{reconstruction of MLS structure from simulated data for a varying number of scans rotated about the surface normal. a) Ground truth. a) 2 angles, b) 3 angles, c ) 6 angles, d) 10 angles. Top right: mean squared error as a function of number of rotation angles.}
\label{fig:MLS_rotation_sim}
\end{figure}

Fig.~\ref{fig:MLS_rotation_sim} shows a reconstruction of a simulated dataset of the Au MLS structure (not the same scale as in Fig.~\ref{fig:combinedscan}) on an Au substrate with a height of $x=5~$nm, with data acquired from a series of rotation angles between 0 and 180 degrees. The rotation angles are linearly spaced between zero and 180 in all cases, and the simulation has settings $\Delta x = 5 ~$nm, $\Delta y = 40 ~$nm, and $\Delta z =700 ~$nm. As can be observed, there is a central ROI where the FOV from all angles overlaps. Due to the anisotropic nature of grazing incidence, the field of view is much larger along the beam direction ($z$) and so only a smaller ROI of the reconstruction will be illuminated by all angles, defined by the maximum $y$ FOV for all angles. This can be observed as an increased degree of "blurring" appearing further away from the ROI. 

The increase in resolution by incorporating multiple rotational angles can be estimated using the relaxed Crowther criterion for multislice ptychography, as outlined in Jacobsen\cite{Jacobsen:18}, where the frequency range of a single slice can be extended beyond the conventional Crowther criterion, and is instead related to the chosen slice thickness $\Delta z$ and D, the maximum size of an object. The number of angles required to achieve the Crowther criterion is calculated to be 28 for reconstruction settings of $\Delta x$ = 5 nm, $\Delta y$ = 40 nm, and $\Delta z$ = 700 nm. A visual representation of the estimated under-sampling can be seen in the supplementary information. Due to the resolution of the slice thickness of the reconstruction process and a large number of transverse pixels, a large amount of Fourier space is sampled even from 10 angles.

\subsection{Simulation: deposited nanoparticles}
To demonstrate the versatility of the reconstruction framework and highlight the applicability to a materials science case, we simulate a random dispersion of nanoparticles deposited on a substrate. This presents a case where it would not be possible to know the prior structure or to provide an initial guess for refinement. 

The particles are circular, plate-like Si nanoparticles with heights of roughly 15-20 nm on an Au substrate, with simulated time-steps where particle growth occurs, to mimic a possible in situ experiment to characterize nanoparticle growth and shape. The only prior knowledge for the reconstruction is that the substrate is Au, the particles must be composed of Si, and the probe and scan positions are known. The reconstruction is initialized as a blank, perfectly flat Au substrate.

The reconstruction involved a total of 360 scans distributed over 8 incidence angles. The heights, radius and position of each particle are all randomized, following a normal distribution. Due to the flexibility of automatic differentiation\cite{Du2021,Jurling2014}, we can apply a constraint where the total mass of particles within the field of view, as inferred through absorption and sample weight measurements, could be calculated, and used to regularize the total mass of the reconstructed volume, however this is not strictly necessary.

\begin{figure}[H]
\centering
\includegraphics[width=\linewidth]{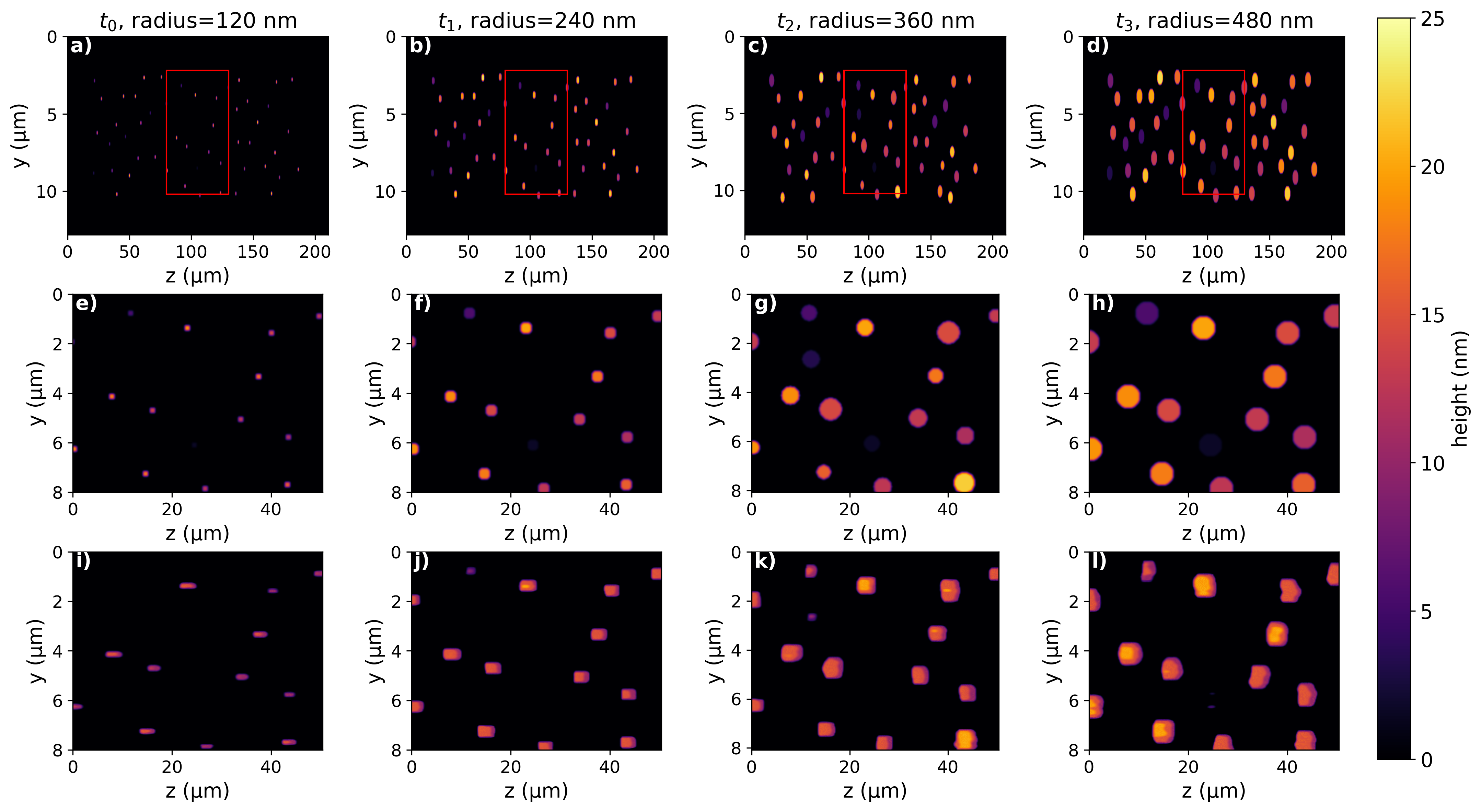}
\caption{Time steps for simulated Si nanoparticle growth on Au substrate. a-d) ground truth. e-h) Ground truth, central ROI. i-l) reconstruction, central ROI.}
\label{fig:nanoparticles_2dfig}
\end{figure}

\begin{table}[H]
    \centering
    \begin{tabular}{c|c|c|c|c}
        \hline
         time point & $t_0$ & $t_1$ & $t_2$ & $t_3$ \\
         \hline
         $\overline{h}_{\text{GT}}~\text{(nm)}$ & 19.4 & 12.7 & 12.7 & 14.1 \\
         \hline
         $\overline{h}_{\text{recon}}~\text{(nm)}$ & 17.9 & 11.7 & 14.0 & 12.7 \\
    \end{tabular}
    \caption{Mean height of nanoparticles for each time step, 2nd row: Ground truth, 3rd row: reconstruction}
    \label{tab:nanoparticles_stats}
\end{table}

Fig.~\ref{fig:nanoparticles_2dfig} shows reconstructions of a distribution of Si nanoparticles on an Au substrate. The $\Delta z$ voxel size is $200~$nm while $\Delta y = 40~$nm and $\Delta x = 5~$nm. The smallest mean radius at the simulated timestep, $t_0$, is only on the order of 2-3 pixels wide, and so some of the smallest particles are not recovered in the earlier time stages. In line with data treatment as one may proceed with imaging of nanoparticles, we can process the image to extract information about the nanoparticles.

\begin{figure}[H]
\centering
\includegraphics[width=\linewidth]{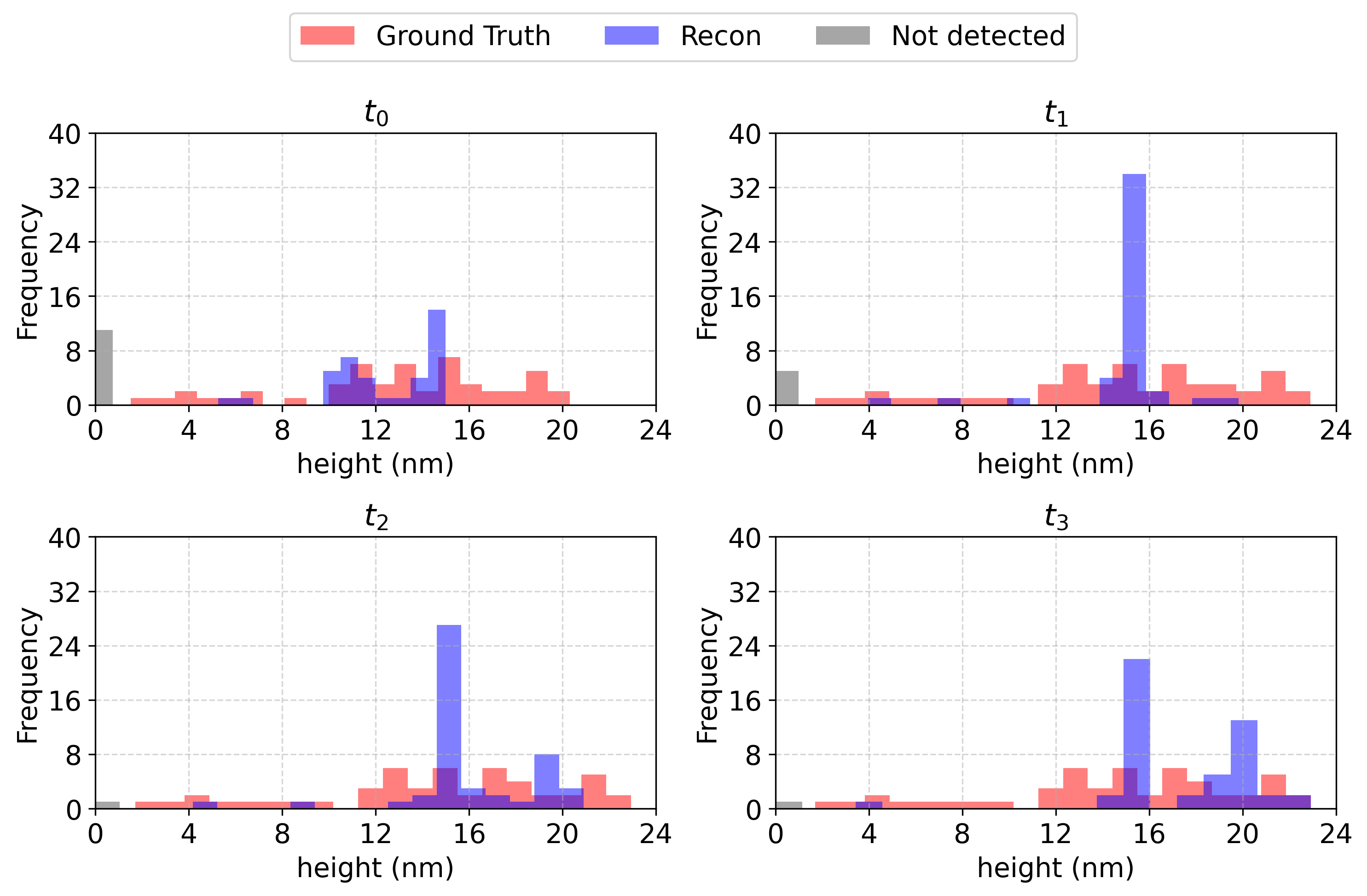}
\caption{Statistics from nanoparticle height fitting. Red: Ground Truth. Blue: reconstruction. Grey bars indicate particles that were not detected at a given $(y,z)$ position from calculations.}
\label{fig:nanoparticles_hist}
\end{figure}

Fig.~\ref{fig:nanoparticles_hist} shows a histogram of the distribution of heights for each time step. In the first two time steps, there are a larger number of reported zero heights as the smallest particles that are smaller than $\delta z$ are not recovered. At later steps, we tend to see that heights tend to converge to a single value, rather than more accurately capture the distribution, i.e. the width of the recovered distribution is underestimated. We attribute this due to the discretization of the simulation grid, as time steps $t_1$ and $t_2$ have the most common calculated heights at h= 15 nm, which corresponds to 3 voxels with given $\Delta x = 5$~nm, suggesting that the framework tends to converge on fixed voxel values over the correct height. The calculated mean particle height for each time step are within 2 nm of the ground truth (Table \ref{tab:nanoparticles_stats}).  

%We can see simulation results to guide scan parameter options - for scans distributed over difference incidence angles, we can see an improvement in recovered heights as opposed to the same number of scans distributed over fewer angles.  

\subsection{Simulation: Multilayer thin film}

\begin{figure}[H]
\centering
\includegraphics[width=\linewidth]{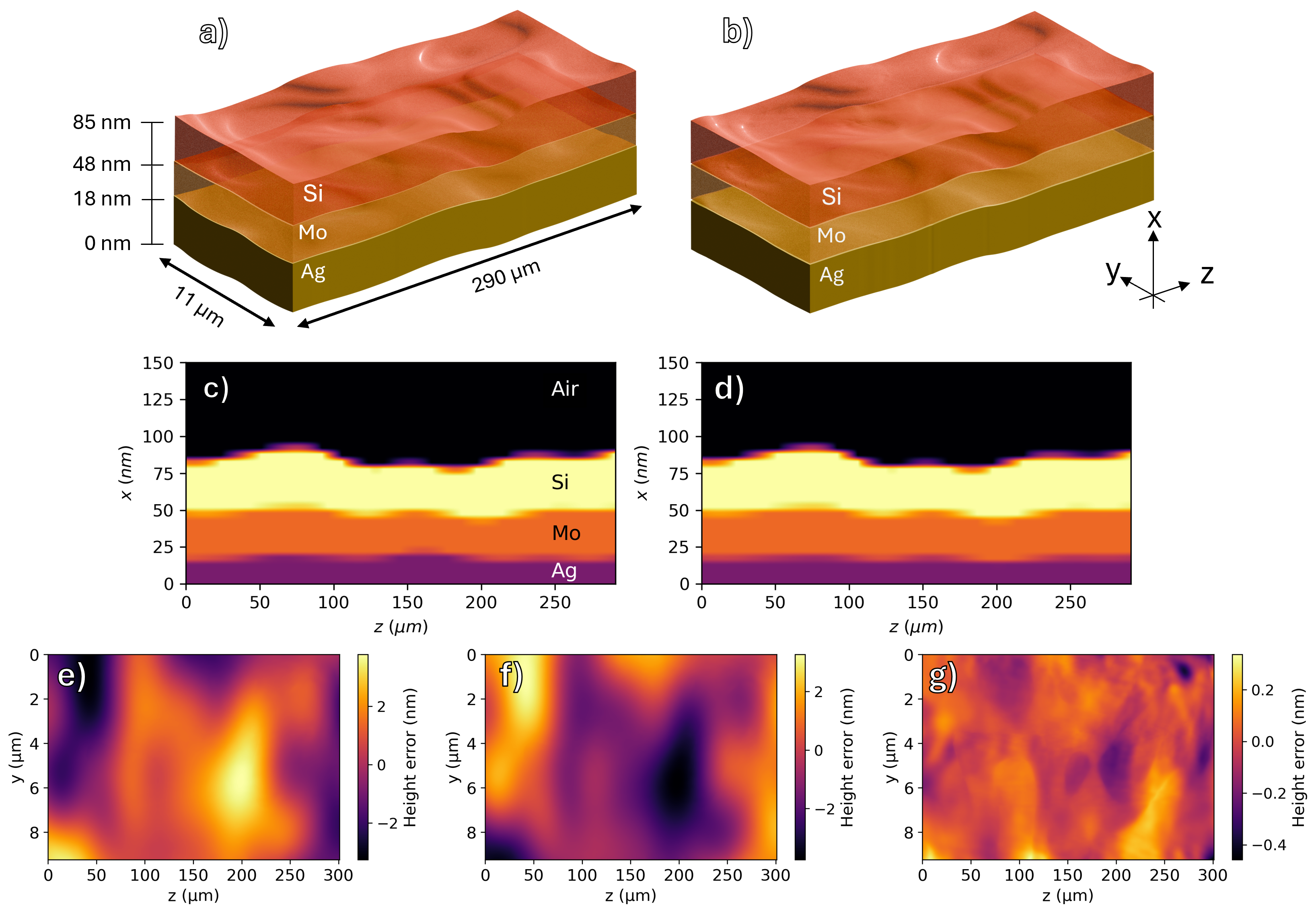}
\caption{Reconstructions of multilayer structure. a) Ground Truth. b) Reconstruction. c) side view of ground truth. d) side view of reconstruction. e) Difference map (error in nm) of Ag layer. f) Difference map (error in nm) of Mo layer. g) Difference map (error in nm) of Si layer. }
\label{fig:multilayer}
\end{figure}

Fig.~\ref{fig:multilayer} shows the ground truth object and reconstruction for a multi-layer, multi-material thin film. This highlights the flexibility of the reconstruction framework for reconstruction of multiple layers with different compositions. The structure consists of an Au substrate, with a deposited Ag layer, followed by an Mo and then Si layer. Fig.~\ref{fig:multilayer}~a) shows the ground truth, whereas Fig.~\ref{fig:multilayer}~b) shows the reconstructed volume. The initial guess used for this reconstruction is composed of a 3-layer structure where the thickness is taken as the mean height of each layer, and each layer is perfectly flat in its initial state. In this case, a stronger prior is used of the average heights, but nothing is known about the specific morphology of the structure beyond this estimation. The layers have an average roughness of $\pm2.3$~nm. The reconstruction is constrained such that the known order of the layers must be preserved (i.e. the volume must be constructed, from substrate upwards, with a layer of Ag, followed by a layer of Mo, then Si). The reconstruction involved a total of 120 scans distributed over 8 incidence angles (from $\theta_i =$ 0.25 to 0.8). The error tends to increase for each layer buried further underneath the surface. The maximum errors for each layer are Si = 0.7~nm, Mo = 2.81~nm, and Ag = 2.84~nm. The structure is built from a series of height maps for each element that follow the optimization specified in eqn.~\ref{equation:wave_interact_partial_voxel}, and may vary independently.

\section{Discussion}

\subsection{Resolution Considerations}
It is important to consider that partial voxel filling with an anisotropic voxel size can have a significant effect on the accurate determination of edges. For example, in Fig.~\ref{fig:combinedscan}, $\Delta x$ = 5~nm whereas $\Delta y$ = 40~nm. The implementation of partial voxel filling as described in  eqn.~\ref{equation:wave_interact_partial_voxel} considers two parts of a partial voxel to interact with the beam simultaneously, without any way of distinguishing the orientation in the plane normal to the beam propagation direction (i.e. whether the components within a partial voxel are stacked vertically along $\Delta x$ or horizontally along $\Delta y$).  In the case of a structure with a 20~nm height, which is smaller than $\Delta y$, heights at the edge of a structure may be poorly estimated. We may also attribute this to why the solved heights do not accurately match the height measured by AFM or by phase shift calculations. The chosen voxel height of $\Delta x$ = 5~nm was chosen to achieve the finest resolution normal to the surface while fitting within memory requirements, but this limits the achievable precision of height compared to previous work \cite{Jorgensen:24} where the height error can be considered to be less than 1 nm. Increasing resolution along $\Delta x$ also means increased memory requirements, but is a promising outlook for future work as higher memory hardware becomes available.

\subsection{Limitations of the current framework}
The largest restriction of the reconstructions presented within this framework are the assumption of a binary system in the case of Fig.~\ref{fig:combinedscan}, and a constraint of a solid continuation of the structure from the substrate (i.e. the sample must be described as a height map, or combination of height maps, stacked vertically from the substrate upwards) with a fixed, known refractive index. This may be challenged when the structure cannot be accurately described having a continuous mixture of two materials from the substrate upwards. In such a case, the optimized volume cannot be represented as an extrapolated height map, and a more complex forward model with a larger parameter space must be used. For example, the structure could be solved as a free volume with freely varying refractive indices, as the framework allows for this, which increases the solution space, decreasing the likelihood of a successful convergence as the reconstruction is prone to converging on local minima and phase wrapping. While this was investigated, it showed that the volume tends to reproduce phase oscillations present in the probe, even when sparsity is enforced with total variation. Therefore such a free optimization may require other constraints or corrections to avoid these phase oscillations, but this remains as future work. Nevertheless the framework is capable of solving a variety of structures with a variety of materials, as is shown in examples of Fig.~\ref{fig:multilayer} and Fig.~\ref{fig:nanoparticles_2dfig}.

\subsection{Ambiguity due to phase wrapping}
As is often the case with optimization methods involving stochastic gradient descent, reconstructions are prone to converging on local minima. Therefore this framework has significant difficulty faithfully reproducing structures with a large $x$ dimension (in this case, referring to a volume size greater than approximately 50 along $x$, or equivalently, a volume size in $x$ of hundreds of nanometers).

Since we maintain the flexibility of this method is to be able to reconstruct surface structures with no initial guess other than a substrate and sample composition, larger and more complicated structures may require further constraints or more complicated initial guesses to avoid converging on local minima. While this is partially addressed by the redundancy gained from overlap in beam footprints, as well as material constraints given by specifying refractive indices, additional constraints or information (such as additional incidence angles) are also needed to accurately resolve the height of these structures encoded in phase shifts. 
\subsection{Hardware and Memory limitations}
The reconstructions performed in this work were all performed on a single Nvidia GeForce RTX 3090 GPU with 24 GB VRAM size. Although the modest hardware requirements allow for a wider use of the reconstruction framework where more powerful hardware may not be available, this places limits on the the total size of the volume, and therefore achievable resolution, for the reconstruction. However, the framework is highly scalable, provided sufficient memory is available to store all gradients, and forward scans can be parallelized as gradients are accumulated from all scans before optimization step occurs. Due to memory limitations, gradients are calculated sequentially for each scan, increasing the computational time significantly.   % \item The voxel size can be further decreased if the memory allows the volume size. We are working with the limitations of a single GPU, 

\section{Conclusion}
We have developed a versatile framework for ptychographic reconstruction which allows new imaging geometries for ptychography performed at grazing incidence, such as a reconstruction involving multiple incidence angles, and rotation about the surface normal. We have also demonstrated a use case for the reconstruction where an accurate initial guess of the structure could not be known, in the case of a distribution of nanoparticles on a surface. We believe the framework could draw many applications in the reconstructions of surface structures and present new avenues for detailed characterization of surface structures in full 3D.

\section{Acknowledgements}
The authors wish to acknowledge the help in SEM image acquisition from Zhongtao Ma and Magnus Björnsson. We acknowledge help from Abdelouadoud Mammeri in AFM measurements. We thank Mirko Holler from the Paul Scherrer Institute for his help during experiments at the cSAXS beamline.  We acknowledge Kaye Morgan and David Paganin for stimulating discussions on mulitslicing and coherence and critical reading of the manuscript. We also wish to acknowledge Chris Jacobsen, Ming Du and Ash Tripati who have provided helpful discussions with computational implementation.

\section{Funding}
Villum Fonden (35997); H2020 Marie Skłodowska-Curie Actions (765604).

\bibliography{sample}

\end{document}